\documentclass[preprint,aps,prd,showpacs,groupedaddress,floatfix,nofootinbib]{revtex4}

\usepackage{dcolumn}% Align table columns on decimal point
\usepackage{graphicx}
\usepackage{epsfig}
\usepackage{hyperref}
\usepackage{graphicx}% Include figure files
\usepackage{dcolumn}% Align table columns on decimal point
\usepackage{amssymb,eucal}
\usepackage{longtable}
\usepackage{amsmath,amssymb,bm}

\newcommand{\U}{\text{U}}
\newcommand{\mathd}{\mathrm{d}}
\newcommand{\mathi}{\mathrm{i}}

\def\nnb{\nonumber}

\newcommand\Tr{{\rm Tr }}

\newcommand\tr{{\rm tr }}

\newcommand{\del}{\partial}

\newcommand{\nn}{\nonumber}

\def\tr{\mathop{\rm tr}\nolimits}
\def\Tr{\mathop{\rm Tr}\nolimits}

\newcommand{\cO}{{\mathcal O}}
\newcommand{\cA}{{\mathcal A}}
\newcommand{\cF}{{\mathcal F}}

\newcommand{\cD}{{\mathcal D}}

\newcommand{\cJ}{{\mathcal J}}

\begin{document}
\preprint{\hfill BARI-TH/2013-676, \\
          \phantom{ } \hfill FTUAM-13-13, \\
          \phantom{ } \hfill IFT-UAM/CSIC-13-070}
\title{Large-distance properties of holographic baryons}
\author{Pietro Colangelo$^a$\footnote{Email: \textsf{Pietro.Colangelo@ba.infn.it}},
Juan Jose Sanz-Cillero$^b$\footnote{Email: \textsf{juanj.sanz@uam.es}},
Fen Zuo$^a$\footnote{Email: \textsf{Fen.Zuo@ba.infn.it}}}
\affiliation{
$^a$Istituto Nazionale di Fisica Nucleare, Sezione di Bari, Italy\\
$^b$Departamento de F\'\i sica Te\'orica and Instituto de F\'\i sica Te\'orica, IFT-UAM/CSIC, Universidad Aut\'onoma de Madrid, Cantoblanco, 28049 Madrid, Spain}
\begin{abstract}
Employing the asymptotic instanton solution in an arbitrary background  of a set of holographic QCD models, we show that  baryon form factors have a precise large-distance behaviour regardless of the background. The dependence coincides with that obtained  from  general chiral soliton models and  large-$N_C$ chiral perturbation theory. The nonlinear terms in the equations of motion are  necessary to recover the correct results. We also show that the holographic currents have the right structure at low energy if the solutions of  the full equation of motion, instead of the linearized ones, are used. The indication is that in this holographic approach, the linearized approximation used in the meson sector is not appropriate for the solitonic description of the baryons.
\end{abstract}

\maketitle

\section{Introduction}
In recent years,  progress has been made in understanding  aspects of QCD employing the duality between a gravity theory in a 5-dimensional space-time and a strong coupled gauge theory on the boundary, the so-called gauge/gravity duality~\cite{Maldacena:1997re,Gubser:1998bc,Witten:1998qj}. Generalizations of the original duality involving ${\cal N}=4$ super Yang-Mills theory
at large $N_C$ towards QCD have been made in many directions. A class of models, remarkably successful in hadron physics, was constructed in ref.~\cite{Son:2003et} using  early ideas of dimensional deconstruction and hidden local symmetry. An explicit string construction of this kind of models was the Sakai-Sugimoto model~\cite{Sakai:2004cn}, obtained embedding $D8/\overline{D8}$ flavor branes into the $D4$ background~\cite{Witten:1998zw}. Models in this class describe very well many properties of hadron physics~\cite{Son:2003et,Sakai:2004cn,Hirn:2005nr,Sakai:2005yt}; in particular,
%due to the spirit of non-linear realization of chiral symmetry,
they reproduce the chiral perturbation theory~($\chi PT$) Lagrangian automatically, with reasonable values of the low-energy constants~(LECs) up to ${\cal O} (p^6)$~\cite{Hirn:2005nr,Sakai:2005yt,Colangelo:2012ipa}.
Moreover, various relations, among different LECs~\cite{Colangelo:2012ipa} and between different amplitudes~\cite{Son:2010vc,Colangelo:2012ipa,Colangelo:2013cxa}, are found in this class of models. Although such relations have not yet been proved exactly, numerical results show that they are useful estimates at large $N_C$. This confirms the interest for this class of models, that we now apply for an analysis of baryons.

The description of baryons as solitonic objects in the mesonic theory started with Skyrme's early work~\cite{Skyrme:1961vq,Skyrme:1961vr,Skyrme:1962vh}, later justified through the large $N_C$ expansion~\cite{'tHooft:1973jz,Witten:1979kh}. Gauge/gravity duality identifies baryons as D-branes wrapping some compact extra dimensions~\cite{Witten:1998xy}, e.g., D4 branes wrapping a sphere $S^4$ in the D4 background~\cite{Sakai:2004cn}. Actually, the wrapping branes can be described as the instanton configuration of resulting effective action of the flavor branes~\cite{Hata:2007mb,Hashimoto:2008zw}. Equivalently, baryons can be considered as  generalized Skyrmions with the inclusion of an infinite number of vector/axial-vector resonant excitations~\cite{Pomarol:2008aa}. Due to the inclusion of the resonance fluctuations, together with the presence of the Chern-Simons~(CS) term, significant improvements in the description  of  baryon properties have been achieved~\cite{Hata:2007mb,Pomarol:2008aa,Hashimoto:2008zw,Panico:2008it} with respect to the  results in the Skyrme model~\cite{Adkins:1983ya}.
However, it has been pointed out that the instanton description in the Sakai-Sugimoto model fails to reproduce a few  model-independent relations for the baryon form factors at large distance, while the treatment in refs.~\cite{Pomarol:2008aa,Panico:2008it} in an analogous model reproduces the expected asymptotic behavior~\cite{Cherman:2009gb}. This would be puzzling, since both models actually belong to the same class  \cite{Becciolini:2009fu,Son:2010vc,Colangelo:2012ipa}.
%\cite{Becciolini:2009fu,Cappiello:2010uy,Son:2010vc,Colangelo:2012ipa}.
 It was argued that the failure to reproduce the correct infrared behavior in the Sakai-Sugimoto model can be attributed to the linear approximation taken there~\cite{Cherman:2011ve}. Indeed, renouncing this approximation and solving asymptotically the full instanton equations, the expected large-distance behavior is recovered~\cite{Cherman:2011ve}.

However, it is still not clear if in all the models of this class, the same asymptotic behaviour of the  form factors can be obtained. In ref.~\cite{Zuo:2011vh} this was claimed to be true  on the basis of  two features of the instanton description: 1) the 5-dimensional Yang-Mills~(YM) plus Chern-Simons action reduces exactly to the Skyrme action when truncated to the pion sector, and 2) the Skyrmion generated from the flat-space instanton preserves the right infrared behavior. Here we  explicitly prove this statement, generalizing the asymptotic instanton solutions found in~\cite{Panico:2008it,Cherman:2011ve} to an arbitrary background. A few ambiguities in the solutions are clarified during our analysis. We also find that, as in the mesonic sector, model-independent results exist in the baryonic sector. This represents a continuation of early investigation of model-independent relations in the Skyrme model~\cite{Adkins:1984cf}.

The paper is organized as follows. In the next section we give a short introduction to the class of holographic models of interest,
and in particular we show that the chiral currents reduce at low energy to those expected from $\chi$PT.
In section \ref{sec:sec3} we obtain the asymptotic instanton solutions in a general background. The asymptotic baryon form factors at large distance are derived in section \ref{sec:sec4}. In the last section  we summarize our results.

\section{Holographic description of baryons}\label{sec:sec2}
\subsection{Action and currents}
We consider a class of holographic models defined by the Yang-Mills  and Chern-Simons  action \cite{Son:2003et,Sakai:2004cn}
\begin{equation}
  S = S_{\rm YM}+S_{\rm CS} ,\label{eq:action}
 \end{equation}
with
\begin{align}
  \label{eq:YM}
  S_{\rm YM} &= -\int\! \mathd^4x \mathd z \tr \left[-f^2(z){\cal F}_{z\mu}^2
  + \frac{1}{2g^2(z)}{\cal F}_{\mu\nu}^2 \right], \\
  \label{eq:CS}
  S_{\rm CS} &=-\frac{ N_c}{24\pi^2} \int\! \tr
  \left[{\cal AF}^2+\frac{i}{2}{\cal A}^3{\cal F}-\frac{1}{10}{\cal A}^5
\right].
\end{align}
The fifth coordinate $z$ runs from $-z_0$ to $z_0$, with $0<z_0\le+\infty$. ${\cal A}(x,z)={\cal A}_\mu dx^\mu+{\cal A}_z dz$, with $\mu=0,1,2,3$, is a 5D $\U(2)$ gauge field, and
${\cal F}=d{\cal A}-i{\cal A} \wedge {\cal A}$ is the field strength.
They are decomposed as
$\displaystyle \cA=\hat A \frac{\mathbf{1}}{2}+A^a\frac{\sigma^a}{2}$ and
$\displaystyle \cF=\hat F \frac{\mathbf{1}}{2}+F^a\frac{\sigma^a}{2}$ in terms of the unit and Pauli matrices.
The functions $f^2(z)$ and $g^2(z)$ are invariant under reflection $z\to -z$,  so that parity can be properly defined in the model. They are  kept arbitrary in our derivation; just for providing the reader with two examples, in the so-called
{``}Cosh'' model they read~\cite{Son:2003et}
\begin{equation}
f^2(z)=\frac{\Lambda^2 \cosh^2(z)}{g_5^2}\,\,\, , ~~g^2(z)=g_5^2\,\,\, ,~~ z_0=\infty,
\end{equation}
while in the
{``}Sakai-Sugimoto" model their expression is~\cite{Sakai:2004cn,Sakai:2005yt}
\begin{equation}
f^2(z)=\frac{\Lambda^2 (1+z^2)}{g_5^2}\,\,\, , ~~g^2(z)=g_5^2 (1+z^2)^{1/3}\,\,\, , ~~z_0=\infty,
\end{equation}
with $\Lambda$ and $g_5$ model parameters.

The CS action is constructed implicitly in the gauge with vanishing gauge potential at the boundaries and, as a result, the boundary term vanishes. In general, the CS action may acquire nonzero boundary terms: for example, in the gauge ${\cal A}_z=0$, additional boundary terms may appear, but the form of the boundary terms are completely fixed by the gauge transformation~\cite{Sakai:2005yt}.  The action (\ref{eq:action}) reduces at low energy to the corresponding terms in $\chi$PT Lagrangian, in particular to the Skyrme term in the even-parity sector and to the Wess-Zumino-Witten~(WZW) term in the odd-parity sector  \cite{Sakai:2005yt,Colangelo:2012ipa}.

Baryons are described by the instanton solution of the equations of motion  derived from the action:
\begin{align}
2\left[
\cD_\nu\bigl(\frac{1}{g^2(z)}\cF^{\nu\mu}\bigr)
-\cD_z\bigl(f^2(z)\cF^{z\mu}\bigr)\right]
-\frac{N_c}{32\pi^2}\epsilon^{\mu NPQR}
\cF_{NP}\cF_{QR}&=0 ,
\label{eq:EOMA_mu}
\\
2\,\cD_\mu\bigl(f^2(z)\cF^{\mu z}\bigr)
+\frac{N_c}{32\pi^2}\epsilon^{\mu\nu\rho\sigma}
\cF_{\mu\nu}\cF_{\rho\sigma}&=0 ,
\label{eq:EOMA_z}
\end{align}
where $\mu,...,\sigma=0,1,2,3$, $N,...,R=\mu,z$,  $\cD_\mu=\partial _\mu-i{\cal A}_\mu$, $\cD_z=\partial _z-i{\cal A}_z$. The 5D anti-symmetric tensor is chosen as $\epsilon^{0123z}=\epsilon^{0123}=1$.

In the meson sector  the so-called bulk-to-boundary propagator is used to calculate correlation functions and form factors. In the baryon sector it is more convenient to express the currents through the on-shell solutions~\cite{Hashimoto:2008zw,Pomarol:2008aa}. Following the idea in ref.~\cite{Son:2003et}, one first introduces the sources as the boundary values of the gauge field:
\begin{equation}
\cA_\mu(x, -z_0)=\ell_\mu(x)\, ,\qquad\qquad
\cA_\mu(x,z_0)=r_\mu(x)\, .
\label{Eq.Amu-bcs}
\end{equation}
The currents are then read from the action through the linear coupling
\begin{equation}
S|_{\mathcal{O}(\ell_\mu,r_\mu)}=-2\int \mathd^4x  \tr(\ell_\mu \cJ_{L}^\mu+r_\mu \cJ_{R}^\mu),\label{eq.current-def}
\end{equation}
with the results~\cite{Hashimoto:2008zw}:
\begin{align}
&\cJ_{L\mu}=
-
\left(f^2(z)\,\cF_{\mu z}\right)\Big |_{z=-z_0}\ ,~~
\cJ_{R\mu}=
\left(f^2(z)\,\cF_{\mu z}\right)\Big |_{z=+z_0}\ .
\label{eq:currentLR}
\end{align}
The vector and axial currents are then given by
\begin{align}
&\cJ_{V\mu}=\cJ_{L\mu}+\cJ_{R\mu}=\Big[\,f^2(z)\cF_{\mu z}\Big]^{z=+z_0}_{z=-z_0}\ ,~~
\cJ_{A\mu}=\cJ_{R\mu}-\cJ_{L\mu}=
\Big[\,\psi_0(z)f^2(z)\cF_{\mu z}\Big]^{z=+z_0}_{z=-z_0}
\ ,
\label{eq:currentVA}
\end{align}
with the decomposition $\displaystyle \cJ=\hat J \frac{\mathbf{1}}{2}+J^a\frac{\sigma^a}{2}$, and $\psi_0(z)$ defined below. The baryon number current, $\displaystyle J_B^\mu=\frac{2}{N_c}\hat J_V^\mu$, ensures through (\ref{eq:EOMA_mu}) that the baryon number is given by the instanton number of the solution~\cite{Hashimoto:2008zw},
\begin{equation}
B=\frac{1}{64\pi^2}\int \mathd ^3x \mathd z ~\epsilon_{mnpq}F^{a}_{mn}F^a_{pq} ,
\end{equation}
where $m,...,q=1,2,3,z$. Notice that the definition (\ref{eq.current-def}) is different from the usual one by a minus sign, thus the nucleons will be described by instantons with $B=-1$.

%We want to clarify some misunderstandings of the currents, especially the large distance behavior.

\subsection{Linear approximation}
The above prescription is quite different from that in the meson sector, where
we obtain the meson fields from the linearized equation of motion
\begin{equation}
\partial_z[f^2(z)\partial_z {\cal A}_\mu(q,z)]=-q^2 {\cal A}_\mu(q,z)/g^2(z) \label{eq:eq}
\end{equation}
which give normalizable solutions $\psi_n(z)$ for the massive resonances~($q^2=m_n^2$) and a massless mode $\psi_0(z)$ corresponding to the pion. In the presence of the chiral sources $\ell_\mu(x)$ and $r_\mu(x)$, the 5D gauge potential ${\cal A}_\mu$ can be expanded as~\cite{Sakai:2005yt,Colangelo:2012ipa}
\begin{equation}
{\cal A}_\mu(x,z)=\ell_\mu(x) \psi_-(z)+r_\mu(x) \psi_+(z)
+\sum_{n=1}^\infty v_\mu^n(x)\psi_{2n-1}(z)
+\sum_{n=1}^\infty a_\mu^n(x)\psi_{2n}(z)\, ,
\label{Eq.Amu-decomposition1}
\end{equation}
with $\psi_\pm(z)=\frac{1}{2}(1\pm\psi_0(z))$.
The fifth component of the gauge potential is related to the pionic field through a Wilson line:
\begin{equation}
U(x)=\exp\{2 \mathi ~\Pi(x) /f_\pi\}=\mbox{P} \exp\left\{i\int^{+z_0}_{-z_0} {\cal A}_z(x,z') dz'\right\}.\label{Eq.U}
\end{equation}
Explicitly, one has
\begin{equation}
{\cal A}_z(x,z)=\Pi(x)\phi_0(z),
\end{equation}
with $\phi_0(z)=\partial_z \psi_0(z)/f_\pi$. Substituting the decomposition into the currents, one obtains the vector and axial-vector currents \cite{Hirn:2005nr,Hashimoto:2008zw}
\begin{align}
&\cJ_{V\mu}
=\sum_{n=1}^\infty
g_{v^n}v_\mu^{n}\ ,~~~
\cJ_{A\mu}
=f_\pi\del_\mu\Pi
+\sum_{n=1}^\infty
g_{a^n}a_\mu^{n}\ ,
\label{eq:currentVA2}
\end{align}
with the couplings
\begin{eqnarray}
g_{v^n}&=&-f^2(z)\partial_z \psi_{2n-1}(z)\mid^{+z_0}_{-z_0}\, ,
%%%\,\,=\,\, \Int_{-z_0}^{z_0} dz\, \Frac{m_{v^n}^2\, \psi_{2n-1}(z) }{g^2(z)}
%%%\nonumber\\
\qquad \qquad g_{a^n} = - f^2(z)\psi_0(z)\partial_z \psi_{2n}(z)\mid^{+z_0}_{-z_0} \, ,
\label{Eq.gv-def}
\end{eqnarray}
and the pion decay constant~\cite{Son:2010vc}
\begin{equation}
f_\pi^2=4\left(\int_{-z_0}^{z_0}\frac{\mathd z}{f^2(z)}\right)^{-1}.
\end{equation}
A generalized vector-meson-dominance~(VMD) structure for the meson system emerges with the obtained expression of the vector current~\cite{Sakai:2005yt}. The two-point correlators can be calculated immediately from the currents (\ref{eq:currentVA2}), and read
\begin{equation}
\Pi_V(-q^2)=\sum_{n=1}^\infty \frac{g_{v^n}^2}{m_{2n-1}^2(-q^2+m_{2n-1}^2)},\quad \Pi_A(-q^2)=-\frac{f_\pi^2}{q^2}+\sum_{n=1}^\infty \frac{g_{a^n}^2}{m_{2n}^2(-q^2+m_{2n}^2)}.
\end{equation}
They have the expected resonance structure, which has also been derived through the bulk-to-boundary propagators~\cite{Son:2010vc,Colangelo:2012ipa}.

In ref.~\cite{Hashimoto:2008zw} this linear approximation was applied to describing the baryons in the Sakai-Sugimoto model, obtaining a similar VMD picture for nucleons. In such an approximation, the static solutions for the resonance fields in (\ref{eq:currentVA2}) are given, after Fourier transforming to the coordinate space, by the Yukawa potential $Y_n(r)$,
\begin{align}
 Y_n(r)=-\frac{1}{4\pi}\frac{e^{-m_n\,r}}{r}\ .
\label{Yukawa}
\end{align}
As a result, the baryon electromagnetic form factors exhibit an exponentially decreasing  behavior at large distance, with no pionic contributions~\cite{Cherman:2009gb}. A manifestation of this fact is that the isovector charge radius is finite in the chiral limit with such an approximation~\cite{Hashimoto:2008zw}. In ref. \cite{Cherman:2011ve} it was shown that, solving the full set of equations of motion, the correct power behavior can be recovered in this model. In the following sections we give a model-independent derivation for the form factors, showing that the linear approximation used in the meson sector is not suitable for describing baryons, and that the resulting VMD picture for  nucleons is doubtful. The key point is that the non-linear terms in the full equations of motion cannot be neglected: these terms make the pion and the resonance  contributions fully entangled, with no way to  clearly separate them.

 Using the linear approximation, also the holographic expressions for the currents (\ref{eq:currentLR}) and (\ref{eq:currentVA}) are doubtful~\cite{Hata:2008xc}. This point can be suitably clarified in the gauge ${\cal A}_z=0$,  where all the elements can be compactly included in the gauge potential~\cite{Hirn:2005nr,Colangelo:2012ipa},
\begin{equation}
{\cal A}_\mu(x,z)=i\Gamma_\mu(x)+\frac{u_\mu(x)}{2}\psi_0(z)
+\sum_{n=1}^\infty \tilde v_\mu^n(x)\psi_{2n-1}(z)
+\sum_{n=1}^\infty \tilde a_\mu^n(x)\psi_{2n}(z)\, ,
\label{Eq.Amu-decomposition2}
\end{equation}
and the commonly used tensors in $\chi$PT~\cite{Bijnens:1999sh,Bijnens:1999hw,Ecker:1988te,Ecker:1989yg},
 $u_\mu(x)$ and $\Gamma_\mu(x)$,  show up naturally:
\begin{eqnarray}
  u_{\mu} \left( x \right)  & \equiv & \mathi \left\{ \xi_R^{\dag}\left( x \right)  \left( \partial_{\mu} - \mathi r_{\mu}
  \right) \xi_R\left( x \right) - \xi_L^{\dag}\left( x \right)  \left( \partial_{\mu} - \mathi \ell_{\mu}
  \right) \xi_L\left( x \right) \right\} \\
    \Gamma_{\mu} \left( x \right) & \equiv & \frac{1}{2}  \left\{ \xi_R^{\dag}\left( x \right)
  \left( \partial_{\mu} - \mathi r_{\mu} \right) \xi_R\left( x \right) + \xi^{\dag}_L\left( x \right)  \left(
  \partial_{\mu} - \mathi \ell_{\mu} \right) \xi_L\left( x \right) \right\}.
\end{eqnarray}
Substituting the decomposition (\ref{Eq.Amu-decomposition2}) in Eq.~(\ref{eq:currentLR}), and throwing away the resonance contributions, the currents reduce to
\begin{equation}
\cJ_{L\mu}=\frac{f_\pi^2}{4} \xi_L u_\mu \xi_L^\dag,~~~~~~\cJ_{R\mu}=-\frac{f_\pi^2}{4} \xi_R u_\mu \xi_R^\dag,
\end{equation}
hence, only the leading term in $\chi$PT, the non-linear sigma term, appears~\cite{Hata:2008xc}.
%Moreover, if one sticks to the linear approximation as in ref.~\cite{Hashimoto:2008zw}, only the term linear to %the pion field, $\pi^a$, remains and the pionic contribution to the vector current vanishes completely. . %Therefore from this point of view, the linear approximation taken there would be doubtful.
How can we reproduce the other terms in the currents, e.g., those related to
the Skyrme and the WZW term? To solve this problem  a different definition of the currents was proposed  \cite{Hata:2008xc}. However, using the full equations of motion it has been shown that the $U(1)$ part of the vector current in Eq.~(\ref{eq:currentVA}) consistently gives the baryon number~\cite{Hashimoto:2008zw}. Here we follow the same idea and show that the currents (\ref{eq:currentLR}) and (\ref{eq:currentVA}) have the right infrared structure if we use the solution to the complete equation (\ref{eq:EOMA_mu}) instead of the linear one (\ref{eq:eq}).

At low energy, the asymptotic solution to the full equation (\ref{eq:EOMA_mu}) can be obtained recursively: one first rewrites (\ref{eq:currentLR}) as integrals over $z$, then replaces the integrands by the remaining terms in Eq.(\ref{eq:EOMA_mu}), and finally substitutes the linear expansion (\ref{Eq.Amu-decomposition2})~\cite{Hashimoto:2008zw}. As a result, the currents  contain the  pionic terms
\begin{eqnarray}
\cJ_{L\mu}&=& \xi_L\left\{\frac{f_\pi^2}{4} u_\mu +\frac{1}{16 e_S^2} [u^\nu,[u_\mu,u_\nu]]\right.\nn\\
&&\left. +\frac{\mathi}{2} L_9 \nabla^\nu [u_\mu,u_\nu] -\frac{1}{4} L_9 [u^\nu,[u_\mu,u_\nu]]-\frac{\mathi N_C}{96\pi^2}\epsilon_{\mu\nu\alpha\beta} u^\nu u^\alpha u^\beta  \right\} \xi_L^\dag , \nn\\
\cJ_{R\mu}&=& \xi_R\left\{-\frac{f_\pi^2}{4} u_\mu -\frac{1}{16 e_S^2} [u^\nu,[u_\mu,u_\nu]]\right.\nn\\
&&\left. +\frac{\mathi}{2} L_9 \nabla^\nu [u_\mu,u_\nu] +\frac{1}{4} L_9 [u^\nu,[u_\mu,u_\nu]]-\frac{\mathi N_C}{96\pi^2}\epsilon_{\mu\nu\alpha\beta} u^\nu u^\alpha u^\beta  \right\} \xi_R^\dag . \label{currents}
\end{eqnarray}
Higher power terms can be  obtained recursively. The covariant derivative $\nabla$ is defined with respect to the vector connection $\Gamma_\mu$:  $\nabla_\mu X \equiv \partial_\mu X +[\Gamma_\mu,X]$.
The Skyrme parameter $e_S$ and the coupling $L_9$ are given by~\cite{Hirn:2005nr,Sakai:2005yt,Colangelo:2012ipa}
\begin{eqnarray}
&&e_S^{-2}=32~L_1=16~ L_2=-\frac{16}{3}L_3=\int_{-z_0}^{z_0} \frac{(1-\psi_0^2)^2}{g^2(z)}  \, \mathd z \, ,\nn\\
&&L_9=-L_{10}=\frac{1}{4}\int_{-z_0}^{z_0} \frac{1-\psi_0^2}{g^2(z)}  \, \mathd z\,.
\end{eqnarray}
In Eqs.(\ref{currents}) we have exactly the currents  obtained from the $\chi$PT Lagrangian  in the chiral limit up to order $\cO(p^4)$. The expected parts corresponding to the nonlinear sigma term, the Skyrme term and the WZW term are present. In addition, one finds two terms
multiplied by the coefficient $L_9$, which are due to the operator $O_9=-i<f_{+\mu\nu}u^\mu u^\nu>$ in the $\cO(p^4)$ chiral Lagrangian with sources, with the covariant tensors $f_{\pm}^{\mu\nu}\equiv
\xi_L ^{-1}\ell^{\mu\nu}\xi_L\pm \xi_R^{-1}r^{\mu\nu}\xi_R$
containing the field-strengths $\ell^{\mu\nu}$ and $r^{\mu\nu}$ of the left and right sources, respectively.
This kind of terms are necessary for the description of the electromagnetic property of pions, and should be included if one wants to describe pions and nucleons in the same model~\cite{Braaten:1986md}. Hence, the nonlinear terms in the equations of motion are necessary to reproduce the corresponding terms in the currents.

 Our conclusion is that  the currents  in Eqs.~(\ref{eq:currentLR}) and (\ref{eq:currentVA}) have the expected structure at low energy provided that one makes use of the solution of the full equation. In addition, as found long ago by Atiyah and Manton~\cite{Atiyah:1989dq}, the Skyrmion solution generated from the flat-space instanton has the $1/r^2$ pion tail at large distance. Based on these two facts, it has been claimed that all the baryonic form factors have the correct infrared behavior in this class of holographic models \cite{Zuo:2011vh}. Here we demonstrate this statement in the following sections using the asymptotic instanton solutions.

\section{Asymptotic instanton solutions}\label{sec:sec3}

\subsection{Static solution}
The asymptotic instanton solutions of Eqs.~(\ref{eq:EOMA_mu},\ref{eq:EOMA_z}) are required for studying various form factor  properties.
First,  let us look for the static instanton solution. Following ref.~\cite{Witten:1976ck}, the cylindrical symmetric ansatz can be employed~\cite{Pomarol:2008aa,Cherman:2011ve}:
\begin{align}
\label{eq:StaticAnsatz}
\bar A_{j}^{a} &= -\frac{\phi_{2}+1}{r^{2}} \epsilon_{j a k} x_{k} + \frac{\phi_{1}}{r^{3}} [\delta_{ja} r^{2} - x_{j} x_{a}]+A_{r} \frac{x_{j}x_{a}}{r^{2}}, \nonumber \\
\bar A_{z}^{a} &= A_{z} \frac{x^{a}}{r},  \\
 \hat {\bar{A}}_{0} &= s, \nonumber
\end{align}
with $\phi_1$, $\phi_2$, $A_r$, $A_z$ and $s$ functions of $z$ and $r$. As for the time-independent solutions, $\bar A^a_0$ vanishes, as well as the components $\hat {\bar A} _j$ are $\hat {\bar A}_z$. With these fields the static mass energy can be expressed as
\begin{align}
\label{eq:StaticSolitonEnergy}
M=M_{YM}+M_{CS},
\end{align}
in terms of
\begin{align}
M_{YM} = 16 \pi \int_{0}^{\infty}dr \int_{-z_0}^{z_0} dz \,  &\left[  \frac{1}{g^2(z)} |D_{r} \phi|^{2} + f^2(z) |D_{z} \phi|^{2}  +\frac{r^{2}}{4g^2(z)}  F_{\alpha\beta}^{2} \right.  \nonumber \\
&\left. +\frac{1}{2g^2(z)r^{2}}   (1-|\phi|^{2})^{2} -\frac{1}{2}r^{2} \left(\frac{1}{g^2(z)} (\partial_{r} s)^{2} + f^2(z) (\partial_{z} s)^{2}\right) \right]
\end{align}
and
\begin{align}
M_{CS} =16\pi  \gamma \int_{0}^{\infty}{dr \int_{-z_0}^{z_0}  dz \, s\,\epsilon^{\alpha\beta}\left[ \partial_{\alpha}(-i\phi^{*}D_{\beta}\phi + h.c) +F_{\alpha\beta}\right]} .
\end{align}
In the above equations we have  $F_{\alpha\beta}=\partial_\alpha A_\beta-\partial_\beta A_\alpha$ with $\alpha, \beta=r,~z$, $\phi=\phi_1+i\phi_2$, $\gamma = N_{c}/(16\pi^{2})$, and the 2D anti-symmetric tensor  chosen as $\epsilon^{rz}=-\epsilon^{zr}=1$. Both $M_{YM}$ and $M_{CS}$ scale as $N_C$, because $f^2(z)$ and $g^{-2}(z)$ contain a factor of $N_C$ as the constant $\gamma$~\cite{Colangelo:2013cxa}. The covariant derivative is defined with respect to $A_\alpha$:  $D_{\alpha} \phi = \partial_{\alpha}\phi -  iA_{\alpha} \phi$. The instanton number reduces to
\begin{equation}
B=\frac{1}{4\pi} \int_{0}^{\infty}{dr \int_{-z_0}^{z_0}  dz \, \,\epsilon^{\alpha\beta}\left[ \partial_{\alpha}(-i\phi^{*}D_{\beta}\phi + h.c) +F_{\alpha\beta}\right]}.
\end{equation}
 From the CS term one reads the coupling between the baryonic current and the U(1) part.

The equations of motion follow from minimizing the static energy:
\begin{align}
\label{eq:2DEoMZcoord}
& D_{r} \left(\frac{1}{g^2(z)} D_{r} \phi \right)+D_{z} \left(f^2(z) D_{z} \phi \right) + \frac{1}{g^2(z) r^2} \phi(1-|\phi|^{2}) -i \gamma \epsilon^{\alpha \beta} \partial_{\alpha} s  D_{\beta} \phi =0, \\
&\partial_{r} \left(r^{2} f^2(z) F_{r z}\right) - f^2(z) \left(i \phi^{\ast} D_{z}\phi + h.c. \right) - \gamma \epsilon^{r z} \partial_{r}s(1-|\phi|^{2}) =0, \\
&\partial_{z} \left(r^{2} f^2(z) F_{z r}\right) - \frac{1}{g^2(z)} \left(i \phi^{\ast} D_{r}\phi + h.c. \right) - \gamma \epsilon^{z r} \partial_{z}s(1-|\phi|^{2}) =0, \\
&\partial_{r} \left( \frac{r^2}{g^2(z)}  \partial_{r} s \right) +\partial_{z} \left( f^2(z) r^{2} \partial_{z} s \right)  + \gamma \epsilon^{\alpha \beta} \left[ \partial_{\alpha}(-i\phi^{\ast}D_{\beta}\phi + h.c.) +F_{\alpha \beta}\right]=0  .
\end{align}
Suitable boundary conditions to solve these equations are needed.  For the two dimensional system the choice of the Lorentz gauge $\partial_\alpha A^\alpha=0$ is convenient \cite{Witten:1976ck}. Moreover, we require all the components except $A_z$ in the ansatz (\ref{eq:StaticAnsatz}) vanish at the boundary,  $r\to \infty$ or $z\to\pm z_0$, to keep the action finite. For $A_z$, a consistent boundary condition with the Lorentz gauge is chosen as $\partial_z A_z=0$. The boundary condition at the point $r=0$ is fixed by requiring $B=-1$, which is not relevant for our asymptotic solutions. As for the infrared point $z=0$, the boundary conditions are automatically satisfied due to the parity of the various fields: $\phi_1$ and $A_r$ are odd under $z\to -z$, while $\phi_2$, $s$ and $A_z$ are even functions of $z$.

The instanton solutions at large distance $r$ can be found term by term in the expansion of $1/r$. The asymptotic ($r\to \infty$) solutions are given by:
{\allowdisplaybreaks
 \begin{align}
\phi_{1} &= \frac{\beta [z - z_0 \psi_0(z)]}{r^{2}} - \frac{\beta z (z^2-z_0^2)-6\beta [z_0 \sigma (z) - z \sigma (z_0)]}{r^4} +\mathcal{O}(1/r^6)\nnb ,\\
\phi_{2} &= -1+\frac{\beta^2(z^2+z_0^2)-2\beta^2zz_0\psi_0(z)}{2r^4}+\mathcal{O}(1/r^6)  \nnb, \\
A_{z} &= \frac{\beta }{r^2}+\frac{\beta(z_0^2-3z^2)+6 \beta[z_0 \omega(z)-\chi(z_0)]}{r^4}+\mathcal{O}(1/r^6) , \label{eq:StaticInstanton1} \\
A_{r} &=-\frac{2 \beta [z - z_0 \psi_0(z)]}{r^3}+\frac{4 \beta z (z^2-z_0^2)- 24 \beta [z_0 \sigma (z) - z \sigma (z_0)]}{r^5}+ \mathcal{O}(1/r^7) \nnb \\
s &=  \frac{\gamma \beta ^3  z_0^3}{f_\pi^2}\, \frac{\psi_0^4(z)-6\psi_0^2(z)+5}{r^{9}}+ \mathcal{O}(1/r^{11}) , \nnb
\end{align}
where $\omega(z)=\int_0^z \psi_0(z) \mathd z$ and $ \sigma(z)=\int_0^z \omega(z) \mathd z$. Taking the metric functions $f^2(z)$ and $g^2(z)$, together with $\psi_0(z)$, in the hard-wall model and the Sakai-Sugimoto model~\cite{Sakai:2004cn,Hirn:2005nr,Colangelo:2012ipa}, the previously found solutions in these models~\cite{Panico:2008it,Cherman:2009gb,Cherman:2011ve} are reproduced. Some discussions are in order. First, the solutions (\ref{eq:StaticInstanton1}) are given for general backgrounds, with arbitrary metric functions $f^2(z)$ and $g^2(z)$ and the corresponding values of $z_0$. There is a special case with $z_0=\infty$, which appears in the {``}Cosh"  and in the Sakai-Sugimoto models. In this case the solutions  in Eq.~(\ref{eq:StaticInstanton1}) are not well defined~\cite{Cherman:2011ve}. One can still use the formal solutions for further calculations, keeping $z_0$ finite and sending $z_0$ to infinity only as a final step. It is argued that this formalism is related to the renormalization procedure~\cite{Cherman:2011ve}.  However, the background with infinite $z_0$ is just a special case of this class of models: in fact, for any specifical background we can choose different kind of coordinates, making $z_0$ infinite or not. It is hard to believe that only in some specifical coordinate system renormalization is needed. Thus, the apparent divergence in the solutions (\ref{eq:StaticInstanton1}) when $z_0=\infty$ is a kind of coordinate singularity.

As shown in the appendix of ref.~\cite{Colangelo:2013cxa}, for any backgrounds there is always a special coordinate choice in which $\displaystyle \tilde f^2(y)=\frac{f_\pi^2}{2}$, the coordinate $y$  being chosen as
\begin{equation}
y=\psi_0(z).
\end{equation}
In such a coordinate system the boundary is located at $y=\pm1$.
One finds that, using the $y$ coordinate, the  solutions (\ref{eq:StaticInstanton1})  simplify to
\begin{align}
\phi_{1} &= \mathcal{O}(1/r^6) \nnb ,\\
\phi_{2} &= -1+\frac{\beta^2_0(1-y^2)}{2r^4}+\mathcal{O}(1/r^6) , \nnb\\
A_{z} &= \frac{\beta_0 }{r^2}+\mathcal{O}(1/r^6) , \label{eq:StaticInstanton2}\\
A_{r} &= \mathcal{O}(1/r^7), \nnb \\
s &= \frac{\gamma \beta_0 ^3}{f_\pi^2 }\, \frac{(y^2-1)(y^2-5)}{r^{9}}+\mathcal{O}(1/r^{11}) . \nnb
\end{align}
Now the fields are well defined, and no divergence appears. Although the infrared behavior of some fields is quite different, the results for physical observable remain unchanged, as discussed in the next section.

Concerning the determination of the parameter $\beta$  in (\ref{eq:StaticInstanton1}),  it has been pointed out  that it is related to the axial coupling $g_A$, the value of the axial form factor at zero momentum transfer $g_A(0)$  \cite{Panico:2008it}. The axial form factor $g_A(q^2)$ is defined as from the matrix element
\begin{equation}
\langle N,p| J_\mu^{A,a}(0)|N',p'\rangle =\bar u_N(p) \left(\frac{\tau^a}{2}\right) [\gamma_\mu \gamma_5 g_A(q^2)+q_\mu\gamma_5h_A(q^2)] u_{N'}(p'),
\end{equation}
with $q=p-p'$.
To keep the action finite when $z_0$ is taken to infinity, $\beta$ should vanish as $1/z_0$ \cite{Cherman:2011ve}. This  information is useful: indeed,  the axial coupling can be computed through the asymptotic solutions, as in \cite{Witten:1976ck}. We find  that $g_A$ is  completely determined by $\beta$. The calculation is in the next section,  here we use the same argument as in the Skyrme model to figure out the relation between $\beta$ and $g_A$. First, we construct the chiral field $U$ in the model with Eq.~(\ref{Eq.U}), which then represents the pseudoscalar part of the solution, i.e, the Skyrmion. From the static instanton solution we obtain
\begin{equation}
F(r)\to \beta z_0 /r^2,\,\,\,r\to \infty \,\,\, ,
\end{equation}
 for $F(r)$ in the Skyrmion profile $U=\exp[i\sigma_a x_a F(r)/r]$. As shown in \cite{Witten:1976ck}, the coefficient of this term is related to $g_A$,
\begin{equation}
\beta z_0= \frac{3 g_A}{8 \pi f_\pi^2},\label{eq:gA}
\end{equation}
or, for the solutions in the  $y$ coordinate,
\begin{equation}
 \beta_0 = \frac{3 g_A}{8 \pi f_\pi^2} \,\,\, .\label{eq:gA2}
\end{equation}
This relation has already been implicitly indicated in the hard wall model~\cite{Panico:2008it}.
If $z_0$ is taken to infinity, $\beta$ vanishes as $1/z_0$; hence,  only the combination $\beta z_0$ must appear in physical quantities.

\subsection{Rotation and quantization}
Let us  consider  the rotation of the static solution. It can be expressed in the  form
\begin{align}
A_m&=A(t)\bar A_m A^\dagger(t),~~~~~~~\hat A_0={\hat{\bar A}}_0,\nonumber\\
A_0&=A(t) \bar A_0 A^\dagger(t)+i A(t)\partial_t A^\dagger(t),~~~~~\hat A_m=\hat {\bar {A}}_m,
\end{align}
 together with the following ansatz for the fields which vanish in the static solutions~\cite{Panico:2008it,Cherman:2011ve}:
\begin{align}
\label{eq:TimeDependentAnsatz}
\bar A^{a}_{0} &= k_{b}\left[\chi_{1}\epsilon^{abc}\hat{x}_{c} -\chi_{2} (  \delta^{ab} - \hat{x}^{a} \hat{x}^{b}) \right] + v (\vec{k} \cdot \hat{x}) \hat{x}^{a} ,\\
\hat{\bar A}_{i} &=\frac{\rho}{r} (k_{i} - (\vec{k}\cdot \hat{x})x_{i}) +B_{r}  (\vec{k}\cdot \hat{x}) \hat{x}_{i} + Q \epsilon_{i b c} k^{b}\hat{x}^{c} ,\\
\hat{\bar A}_{z} &= B_{z}  (\vec{k}\cdot \hat{x}),
\end{align}
with $\hat x_i=x_i/r$.
The rotation velocity $K$ is defined as
\begin{equation}
K=k_a \sigma^a/2=-iA^\dagger \mathd A/\mathd t
\end{equation}
and  takes a constant value in the semiclassical approximation.
These fluctuations contribute to the rotational energy of the instanton,
\begin{align}
\label{eq:RotatingSoliton}
\mathcal{L} = -M + \frac{\Lambda}{2} k_{a} k^{a}.
\end{align}
The moment of inertia $\Lambda$ is given by
\begin{align}\label{eq:moment_of_inertia}
\Lambda =&\frac{16\pi }{3}\int_0^\infty dr\int^{-z_0}_{z_0}dz\,
\Big[
-\frac{1}{g^2(z)} (D_r\rho)^2 - f^2(z) (D_z\rho)^2
-\frac{r^2}{g^2(z)}  (\partial_r Q)^2 - r^2 f^2(z) (\partial_z Q)^2
\nonumber \\
& -2\frac{1}{g^(z)} Q^2
-\frac{r^2}{2g^2(z)}  B_{\alpha\beta}^2
+\frac{r^2}{g^2(z)}  |D_r\chi|^2 + r^2 f^2(z) |D_z\chi|^2
+\frac{r^2}{2g^2(z)} (\partial_r v)^2 + \frac{r^2}{2} f^2(z)(\partial_z v)^2
\nonumber \\
&+
\frac{1}{g^2(z)} (|\chi|^2+v^2)(1+|\phi|^2)
-4v\phi_p\chi_p
+\gamma
\Big(
-2\epsilon^{\alpha\beta}D_{\alpha}\rho \, \chi_p (D_{\beta}\phi)_p
\nonumber \\
&+2\epsilon^{\alpha\beta}\partial_{\alpha}(r Q) \chi_{p}\epsilon^{pq}(D_{\beta}\phi)_q
-v(\epsilon^{\alpha\beta}B_{\alpha\beta}(|\phi|^2 -1)/2
+r Q\epsilon^{\alpha\beta}F_{\alpha\beta}
+2rQ\epsilon^{\alpha\beta} D_\alpha \rho\partial_\beta s
) \big)\Big],
\end{align}
where $p,q=1,2$, $\chi=\chi_1+i \chi_2$, $D_{\alpha} \chi = \partial_{\alpha}\chi -  iA_{\alpha} \chi$, $D_{\alpha} \rho = \partial_{\alpha}\rho -  B_{\alpha}$, and $B_{\alpha\beta}=\partial_\alpha B_\beta-\partial_\beta B_\alpha$ is the field strength of $B_\alpha$. Although the explicit form of the moment  of inertia (\ref{eq:moment_of_inertia}) is quite involved, the structure of the Lagrangian (\ref{eq:RotatingSoliton}) is  the same as that in the Skyrme model~\cite{Adkins:1983ya}. The masses of different baryonic states are then given by
\begin{equation}
E=M+\frac{1}{2\Lambda}J(J+1).
\end{equation}
An immediate consequence is that the nucleon-delta mass splitting is $\displaystyle \Delta=\frac{3}{2\Lambda}$. Since $\Lambda$ scales the same as $M$ at large $N_C$, the mass splitting is of order $1/N_C$.
The equations can be obtained by minimizing the energy:
{\allowdisplaybreaks
\begin{align}\label{eq:collective_eom}
& \frac{1}{r^2}\partial_r (r^2 \partial_r v) + g^2(z) \partial_z( f^2(z) \partial_z v) - \frac{2}{r^2} ( v(1+|\phi|^2)- \chi \phi^\dagger -\phi \chi^\dagger )\nonumber\\
& \qquad\qquad\qquad\qquad\qquad\qquad\qquad\qquad\qquad+ \frac{\gamma g^2(z)}{r^2 }[(|\phi|^2-1) B_{rz} + 2rQF_{rz}]=0, \nonumber \\
& \frac{1}{r^2} D_r (r^2 D_r \chi) + g^2(z) D_z (f^2(z) D_z \chi) + \frac{1}{r^2} (2v\phi-(1+|\phi|^2)\chi)\nonumber\\
& \qquad\qquad\qquad\qquad\qquad\qquad\qquad\qquad\qquad- \frac{\gamma g^2(z)}{r^2} \epsilon^{\alpha\beta}(D_\alpha \phi (i \partial_\beta Q +D_\beta \rho)=0, \nonumber \\
& \frac{1}{r^2}\partial_r (r^2 \partial_r Q) + g^2(z) \partial_z (f^2(z)\partial_z Q) - \frac{2}{r^2} Q \nonumber \\
& \qquad- \frac{\gamma g^2(z)}{ 2r } \epsilon^{\alpha\beta}[ (i D_\alpha \phi(D_\beta \chi)^\dagger + h.c. )+ F_{\alpha\beta} (2v-\chi\phi^\dagger -\phi\chi^\dagger)/2 -2 D_\alpha\rho \, \partial_\beta s]=0, \nonumber \\
& \partial_r (D_r \rho) + g^2(z) \partial_z (f^2(z) D_z \rho)
-\frac{\gamma g^2(z)}{2 } \epsilon^{\alpha\beta} [(D_\alpha \phi (D_\beta \chi)^\dagger + h.c.) \nonumber\\
& \qquad\qquad\qquad\qquad\qquad\qquad\qquad\qquad\qquad+ i F_{\alpha\beta}(\phi\chi^\dagger-\chi\phi^\dagger)/2+2\partial_\alpha(rQ)\partial_\beta s ]=0, \nonumber \\
& g^2(z) \partial_z(f^2 (z) B_{zr}) + \frac{2}{r^2}D_r \rho
+\frac{\gamma g^2(z)}{r^2 } [ ((\chi-v\phi)(D_z \phi)^\dagger + h.c.) + (1-|\phi|^2) \partial_z v - 2r Q\partial_z s]=0, \nonumber \\
& \frac{1}{r^2}\partial_r(r^2 B_{rz}) + \frac{2}{r^2}D_z \rho
-\frac{\gamma}{r^2 f^2 (z)} [ ((\chi-v\phi)(D_r \phi)^\dagger + h.c.) + (1-|\phi|^2) \partial_r v - 2r Q\partial_r s]=0.
\end{align} }
Choosing the 2D Lorentz gauge for $B_\alpha$~\cite{Panico:2008it},  the boundary conditions of the component fields are completely determined by the static solution, together with the properties under $z\to -z$. The corresponding solutions read:
\begin{align}
\label{eq:TimeDependentSolutions}
\chi_{1} &=\frac{\beta [z - z_0 \psi_0(z)]}{r^{2}} - \frac{\beta z (z^2-z_0^2)-6\beta [z_0 \sigma (z) - z \sigma (z_0)]}{r^4}+\mathcal{O}(1/r^6) \nnb ,\\
\chi_{2} &=  -1+\frac{\beta^2(z^2+z_0^2)-2\beta^2zz_0\psi_0(z)}{2r^4}+\mathcal{O}(1/r^6)  \nnb, \\
%v &= -1 + \frac{2\beta^2z_0^2}{f_\pi^2 r^6}\left\{\left[1-\psi_0(z)\right]\int_0^z \frac{1-\psi_0^2(z)}{g^2(z)}\mathd z +\int_z^{z_0} %\frac{[1-\psi_0(z)][1-\psi_0^2(z)]}{g^2(z)} \mathd z \right\} ,\nnb \\
v &= 1 - \frac{2\beta^2z_0^2}{f_\pi^2 }\frac{v^{(6)}(z)}{r^6}+\mathcal{O}(1/r^8) ,\nnb\\
Q &= + \frac{\gamma \beta ^3  z_0^3}{f_\pi^2}\, \frac{\psi_0^4(z)-6\psi_0^2(z)+5}{r^{8}}+\mathcal{O}(1/r^{10}),  \\
B_{z} &=\frac{\hat{\beta} }{r^2}+\frac{\hat{\beta}(z_0^2-3z^2)+6 \hat{\beta}[z_0 \omega(z)-\sigma(z_0)]}{r^4} +\mathcal{O}(1/r^6) , \nnb \\
B_{r} &=-\frac{2 \hat{\beta} [z - z_0 \psi_0(z)]}{r^3}+\frac{4 \hat{\beta} z (z^2-z_0^2)- 24 \hat{\beta} [z_0 \sigma (z) - z \sigma (z_0)]}{r^5}+\mathcal{O}(1/r^7)  \nnb \\
\rho &= \frac{\hat{\beta} [z - z_0 \psi_0(z)]}{r^{2}} - \frac{\hat{\beta} z (z^2-z_0^2)-6\hat{\beta} [z_0 \sigma (z) - z \sigma (z_0)]}{r^4}+\mathcal{O}(1/r^6) , \nnb
\end{align}
where
\begin{equation}
v^{(6)}(z)=\left\{\left[1-\psi_0(z)\right]\int_0^z \frac{1-\psi_0^2(z)}{g^2(z)}\mathd z +\int_z^{z_0} \frac{[1-\psi_0(z)][1-\psi_0^2(z)]}{g^2(z)} \mathd z \right\}.
\end{equation}
Up to the  considered order, $\chi$ and $Q$  have the same expressions as $\phi$ and $r\cdot s$, while $B_\alpha$ has the same expressions as $A_\alpha$ with $\beta$ replaced by a new parameter $\hat \beta$. In the $y$ coordinate system the expressions are simplified:
\begin{align}
\label{eq:TimeDependentSolutions}
\chi_{1} &=\mathcal{O}(1/r^6), \nnb \\
\chi_{2} &= -1+\frac{\beta_0^2(1-y^2)}{2r^4}+\mathcal{O}(1/r^6),  \nnb\\
v &= 1 - \frac{2\beta_0^2}{f_\pi^2 }\frac{v^{(6)}(y)}{r^6}+\mathcal{O}(1/r^8),\nnb\\
Q &=  \frac{\gamma \beta_0 ^3}{f_\pi^2 }\,\, \frac{(y^2-1)(y^2-5)}{r^{8}}+\mathcal{O}(1/r^{10}),  \\
B_{z} &=\frac{ {\hat\beta}_0 }{r^2}+\mathcal{O}(1/r^6), \nnb \\
B_{r} &=\mathcal{O}(1/r^7), \nnb \\
\rho &= \mathcal{O}(1/r^6), \nnb
\end{align}
with
\begin{equation}
v^{(6)}(y)=\left\{\left[1-y\right]\int_0^y \frac{1-y'^2}{\tilde g^2(y')}\mathd y' +\int_y^{1} \frac{[1-y'][1-y'^2]}{\tilde g^2(y')} \mathd y' \right\}.
\end{equation}
The parameter $\hat \beta$ is fixed by
\begin{equation}
\hat{\beta}z_0=-\frac{\Lambda \hat{g}_A}{4\pi\hat{f}_\pi^2},\label{eq:gA0}
\end{equation}
where $\hat{f_\pi}$ and $\hat{g}_A$ are the decay constant and the axial coupling in the isoscalar sector.
$\hat{g}_A$  can be defined  as the corresponding coupling  in the SU(2) part, through the matrix elements
 \begin{equation}
\langle N,p| \hat J_\mu^{A}(0)|N',p'\rangle =\bar u_N(p) (\frac{\mathbf{1}}{2}) [\gamma_\mu \gamma_5 \hat g_A(q^2)+q_\mu\gamma_5 \hat h_A(q^2)] u_{N'}(p'),
\end{equation}
with $q=p-p'$, hence $\hat{g}_A=\hat g_A(0)$. Since in the meson part the exact U(2) symmetry holds, the equality $\hat{f_\pi}=f_\pi$ is always valid. Then $\hat{g}_A$ is determined solely by the parameter $\hat{\beta}$ in the rotation solution;  $\hat{\beta}$ should vanish as $1/z_0$ if $z_0= \infty$~\cite{Cherman:2011ve}.

\section{Asymptotic form factors}\label{sec:sec4}
\subsection{Electromagnetic form factors}
The instanton solutions can be used to calculate the nucleon form factors.
The electromagnetic form factors, after Fourier transforming to the coordinates space, are given by
\begin{align}
\label{eq:FormFactorDefinition}
\tilde{G}_E^{I=0}(r) &= \frac{1}{4\pi} \int~d\Omega\langle p\uparrow|\frac{J_B^{0}}{2}|p\uparrow\rangle   \\
\tilde{G}_M^{I=0}(r) &= \frac{1}{4\pi} \int~d\Omega{1\over2}\varepsilon_{i j 3}\langle p \uparrow|x_{i} \frac{J_B^{j}}{2}|p\uparrow\rangle \\
\tilde{G}_E^{I=1}(r) &= \frac{1}{4\pi} \int~d\Omega\langle p\uparrow|J_V^{\mu = 0, a = 3} | p \uparrow\rangle  \\
\tilde{G}_M^{I=1}(r) &= \frac{1}{4\pi} \int~d\Omega{1\over2} \varepsilon_{i j 3}\langle p\uparrow| x_{i} J_V^{\mu=j, a=3} | p \uparrow\rangle,
\end{align}
where $|p\uparrow\rangle$ is a normalized state for a spin up proton.
%with the baryon number current $\displaystyle J_B^\mu=\frac{2}{N_c}\hat J_V^\mu$. %Notice that in our notation, the above %definitions %is the same as in ref.~\cite{Adkins1987},which
%differ from those in ref.~\cite{Cherman:2009gb} by a factor of $2$.
Replacing the soliton ansatz in the definition of the currents, the form factors can be expressed as
\begin{align}
\label{eq:HolographicFormFactors}
\tilde{G}_E^{I=0}(r) &= -{1\over N_c}  \left[ f^2(z)\partial_z s  \right]^{z_0}_{-z_0}, \\
\tilde{G}_M^{I=0}(r)&= -{1\over 6N_c\Lambda}  \left[ r f^2(z)\partial_z Q~ \right]^{z_0}_{-z_0}, \\
\tilde{G}_E^{I=1}(r) &= {1\over 6\Lambda} \left[f^2(z)(\partial_z v-2 (\partial_z\chi_2- A_z \chi_1)) \right]^{z_0}_{-z_0}, \\
\tilde{G}_M^{I=1}(r) &= -{1\over9} \left[ f^2(z)(\partial_z \phi_2 - A_z \phi_1)\right]^{z_0}_{-z_0}.
\end{align}
With the  instanton solutions, we can calculate the form factors  in the large distance region either using  the general solutions or the simplified ones in the $y$ coordinate:
\begin{align}
\label{eq:FormFactorResults}
\tilde{G}_E^{I=0}(r) &\to \frac{8  \gamma \beta ^3 z_0^3  }{N_c} \frac{1}{r^9},  \\
\tilde{G}_M^{I=0}(r) & \to \frac{4  \gamma \beta ^3 z_0^3  }{3 N_c \Lambda }\frac{1}{r^7} , \\
\tilde{G}_E^{I=1}(r) & \to \frac{ f_\pi^2 \beta ^2 z_0^2}{3   \Lambda  }\frac{1}{r^4} ,  \\
\tilde{G}_M^{I=1}(r) & \to \frac{ f_\pi^2 \beta^2  z_0^2}{9 } \frac{1}{r^4}.
\end{align}
These quantities depend on $\beta$ through the combination $\beta z_0$~\cite{Cherman:2011ve}, and  remain finite regardless $z_0$ is infinite or not. The leading behavior
of the isovector form factors is determined by $\chi_2$ and $\phi_2$.
%Notice that the inertial of momentum is related the nucleon-delta mass splitting $\Delta$ as $\Delta=
%\frac{3}{2\Lambda}$, as in the Skyrme model~\cite{Adkins:1983ya,Adkins1987}.
Moreover, using the expression (\ref{eq:gA}) for $\beta z_0$, we may express the asymptotic form factors,
\begin{align}
\label{eq:FF_GSE}
\tilde{G}^{I=0}_{E}(r) &\to \frac{3^3}{2^{10}\pi^5} \frac{1}{f_{\pi}^3}\left(\frac{g_A}{f_{\pi}}\right)^3 \frac{1}{r^9} ,\\
\label{eq:FF_GSM}
\tilde{G}^{I=0}_{M}(r) &\to \frac{3\Delta}{2^{10}  \pi^5}\frac{1}{f_{\pi}^{3}} \left(\frac{g_A}{f_{\pi}}\right)^3 \frac{1}{r^7} ,\\
\label{eq:FF_GVE}
\tilde{G}^{I=1}_{E}(r) &\to \frac{\Delta}{2^5  \pi^2} \left(\frac{g_A}{f_{\pi}}\right)^2 \frac{1}{r^4} , \\
\label{eq:FF_GVM}
\tilde{G}^{I=1}_{M}(r) &\to \frac{1}{2^6 \pi^2} \left(\frac{g_A}{f_{\pi}}\right)^2 \frac{1}{r^4} .
\end{align}
 The same results were found in the Skyrme model and in general chiral soliton models~\cite{Cherman:2009gb}, and were also derived in large $N_c$ $\chi$PT~\cite{Cohen:2012wm}. As a consequence, one finds the ratio \cite{Cherman:2009gb,Cohen:2012wm}
\begin{align}\label{eq:SSmodelRatio}
\lim_{r\to\infty} r^{2} \frac{\tilde{G}_{E}^{I=0}\tilde{G}_{E}^{I=1}}{\tilde{G}_{M}^{I=0}\tilde{G}_{M}^{I=1}} = 18.
\end{align}
 Compared to the results in refs. ~\cite{Cherman:2009gb,Cherman:2011ve}, in the holographic framework we recover  not only the combination (\ref{eq:SSmodelRatio}), but also the asymptotic form of individual form factors. The  asymptotic expressions are independent of the backgrounds, confirming the statement  that they are universal in large $N_C$ and chiral limits  \cite{Cohen:2012wm}.

 It is also possible to derive the next-to-leading terms for all the form factors in the $1/r$ expansion. For the isovector ones, the next-to-leading terms are ${\cal O}(1/r^6)$, with the coefficients proportional to $L_9$. As discussed in section \ref{sec:sec2}, the terms related to ${\cal O}_9$ do not appear in the Noether currents of the Skyrme model. As a result, the next-to-leading terms of the isovector form factors in the Skyrme model are ${\cal O}(1/r^{10})$, with the coefficients related to the Skyrme parameter.

\subsection{Goldberger-Treiman relation and the axial form factor}
As a complement of the above derivation, we compute the axial coupling and the $\pi NN$ vertex. With the ansatz (\ref{eq:StaticAnsatz}) the axial current can be expressed as
\begin{equation}
J_j^{A,a}=-\frac{1}{2} \Tr [A(t)\sigma^b A^\dagger(t) \sigma^a] \left\{\psi_0(z)f^2(z)\left[ \frac{D_z \phi_1}{r}(\delta_{jb}-\hat x_j \hat x_b)-F_{rz}\hat x_j \hat x_b\right]\right\}_{-z_0}^{+z_0},
\end{equation}
where $D_z \phi_1=\partial_z \phi_1+A_z \phi_2$.
Asymptotically, we have
\begin{equation}
J_j^{A,a}\sim \frac{\beta z_0 f_\pi^2}{2r^2}(\delta_{jb}-3\hat x_j \hat x_b) \Tr [A(t)\sigma^b A^\dagger(t) \sigma^a].
\end{equation}
In semiclassical approximation we have the reduced current conservation relation $\partial_i J_i^A=0$~\cite{Adkins:1983ya}, which helps to simplify the integral $\int \mathd ^3x J_i^A$:
\begin{equation}
\int \mathd ^3x J_i^{A,a}=-\frac{4\pi}{3}f_\pi^2\beta z_0 \Tr[A\sigma^i A^\dagger \sigma^a].
\end{equation}
Evaluating the expectation value on the nucleon state, Eq.~(\ref{eq:gA}) is recovered. In the derivation one needs the  relation~\cite{Adkins:1983ya}
\begin{equation}
\langle N|\Tr[A\sigma^i A^\dagger \sigma^a] |N\rangle=-\frac{8}{3}\langle N| S^i I^a |N\rangle,
\end{equation}
with $S$ and $I$ the spin and isospin operators.

To obtain the $\pi NN$ vortex, one needs to extract the pion field of the rotating instanton, for which the chiral field $U$ is
\begin{equation}
U(x^\mu)=\mbox{P} \exp\left\{i\int^{+z_0}_{-z_0} A(t){\bar{\cal A}}_z(x^\mu,z')A^\dagger(t) dz'\right\} .
\end{equation}
This fixes the large distance behavior of the pion field:
\begin{equation}
\pi^a\sim \frac{\beta z_0 f_\pi}{2r^2}\hat x_i \Tr [A(t) \sigma^i A(t)^\dagger \sigma^a].
\end{equation}
Taking the expectation value on a nucleon state, the $\pi NN$ coupling can be computed~\cite{Adkins:1983ya},
\begin{equation}
\langle N| \pi^a |N\rangle\sim-\frac{g_{\pi NN}}{2\pi M_N} \frac{x_i}{r^3}\langle N| S^i I^a |N\rangle.
\end{equation}
Combining with Eq.~(\ref{eq:gA}), the Goldberger-Treiman relation is recovered:
\begin{equation}
\frac{g_{\pi NN}}{M_N}=\frac{g_A}{f_\pi}.
\end{equation}
In the same manner, one can verify Eq.~(\ref{eq:gA0}) finding the Goldberger-Treiman relation in the isoscalar sector.  A similar derivation  has been carried out  in the Sakai-Sugimoto model with the linear approximation \cite{Hashimoto:2008zw} .

The axial coupling is the value of the form factor at zero momentum. One can also study the axial form factor in the coordinate space. Analogously to the electromagnetic form factors, we define
\begin{equation}
\tilde{G}_A(r) = \frac{1}{4\pi} \int~d\Omega\langle p\uparrow|J_A^{j = 3, a = 3} | p \uparrow\rangle  .\label{eq:AFF}
\end{equation}
The leading $1/r^3$ term of the axial current vanishes after the angular integration, hence higher terms in the asymptotic solutions are needed. For simplicity, we  use the $y$ coordinate, with the required fields given by
\begin{align}
\phi_1&=\frac{ \beta_0^3y(y^2-1)}{3r^6}+\mathcal{O}(r^{-8}),\nonumber\\
\phi_2&=-1+\frac{\beta_0^2(1-y^2)}{2r^4}+\mathcal{O}(r^{-6}),\nonumber\\
A_z&=\frac{\beta_0}{r^2}-\frac{\beta_0^3}{21r^6}+\mathcal{O}(r^{-8}),\label{eq:Az}  \\
A_r&=\mathcal{O}(r^{-9}). \nnb
\end{align}
Substituting these into the definition (\ref{eq:AFF}) and using the relation (\ref{eq:gA2}), we find
\begin{equation}
\tilde{G}_A(r)\to \frac{3}{7\cdot 2^6 \pi^3}\frac{1}{f_\pi}\left(\frac{g_A}{f_\pi}\right)^3\frac{1}{r^7}, \label{eq:AFF2}
\end{equation}
with the power implicitly indicated in ref.~\cite{Panico:2008it}. The next-to-leading term can also be obtained, which turns to be ${\cal O}(1/r^9)$. In this case, both the Skyrme term and ${\cal O}_9$ contribute to the next term. It is not difficult to confirm the leading behavior in the Skyrme model. The solution  for $A_z$ in (\ref{eq:Az}) implies that the corresponding solution for the chiral angle in the Skyrme model is
\begin{equation}
F(r)=\frac{\beta_0}{r^2}-\frac{\beta_0^3}{21 r^6}+\mathcal{O}(r^{-8}).
\end{equation}
Indeed one finds this solution from the equation of motion in the original model. Substituting this into the corresponding axial current \cite{Adkins:1983ya,Adkins1987}, one gets the asymptotic form factor (\ref{eq:AFF2}). Parallel derivation in the isoscalar sector is more involved, and currently under investigation.

\section{Conclusions}
We have extended to the baryon sector our analyses of the meson sector in a class of holographic models. We have found model-independent expressions for various baryon form factors at large distance, in agreement with the results of  chiral soliton models and  large $N_C$ $\chi$PT predictions. These results support the validity of the holographic approach,  confirming that it captures key features of QCD in large $N_C$ and chiral limits. Our analysis suggests that only when the full instanton solutions are considered  the correct structure of currents and  form factors is recovered. The obtained asymptotic solutions will be useful to construct full instanton solutions in this class of models.

\section*{Acknowledgments}
FZ thanks Feng-Kun Guo for discussions at the early stage of the work.
This work is partially supported by the Italian Miur PRIN 2009, the MICINN-INFN fund AIC-D-2011-0818, the MICINN, Spain, under contract FPA2010-17747 and  Consolider-Ingenio CPAN CSD2007-00042,  the Comunidad de Madrid through Proyecto HEPHACOS S2009/ESP-1473, the Spanish MINECO Centro de excelencia Severo Ochoa Program under grant SEV-2012-0249, and by the National Natural Science Foundation of China under Grant No. 11135011.

%\bibliography{Mybib}

\begin{thebibliography}{10}

\bibitem{Maldacena:1997re}
J.~M.~Maldacena,
\newblock Adv.~Theor.~Math.~Phys. \textbf{2} (1998) 231
  [\href{http://arxiv.org/abs/hep-th/9711200}{hep-th/9711200}].

\bibitem{Gubser:1998bc}
S.~S.~Gubser, I.~R. Klebanov, and A.~M. ~Polyakov,
\newblock Phys.~Lett. \textbf{B428} (1998) 105
  [\href{http://arxiv.org/abs/hep-th/9802109}{hep-th/9802109}].

\bibitem{Witten:1998qj}
E.~Witten,
\newblock Adv.~Theor.~Math.~Phys. \textbf{2} (1998) 253
  [\href{http://arxiv.org/abs/hep-th/9802150}{hep-th/9802150}].

\bibitem{Son:2003et}
D.~T.~Son and M.~A.~Stephanov,
\newblock Phys.~Rev. \textbf{D69} (2004) 065020
  [\href{http://arxiv.org/abs/hep-ph/0304182}{hep-ph/0304182}].

\bibitem{Sakai:2004cn}
T.~Sakai and S.~Sugimoto,
\newblock Prog.~Theor.~Phys. \textbf{113} (2005) 843
  [\href{http://arxiv.org/abs/hep-th/0412141}{hep-th/0412141}].

\bibitem{Witten:1998zw}
E.~Witten,
\newblock Adv.~Theor.~Math.~Phys. \textbf{2} (1998) 505
  [\href{http://arxiv.org/abs/hep-th/9803131}{hep-th/9803131}].

\bibitem{Hirn:2005nr}
J.~Hirn and V.~Sanz,
\newblock JHEP \textbf{12} (2005) 030
  [\href{http://arxiv.org/abs/hep-ph/0507049}{hep-ph/0507049}].

\bibitem{Sakai:2005yt}
T.~Sakai and S.~Sugimoto,
\newblock Prog.~Theor.~Phys. \textbf{114} (2005) 1083
  [\href{http://arxiv.org/abs/hep-th/0507073}{hep-th/0507073}].

\bibitem{Colangelo:2012ipa}
P.~Colangelo, J.~J.~Sanz-Cillero,  F.~Zuo,
\newblock JHEP \textbf{1211} (2012) 012
  [\href{http://arxiv.org/abs/1207.5744}{arXiv:1207.5744}].

\bibitem{Son:2010vc}
\newblock D.~T. ~Son and N.~Yamamoto (2010)
  [\href{http://arxiv.org/abs/1010.0718}{arXiv:1010.0718}].

\bibitem{Colangelo:2013cxa}
P.~Colangelo, J.~J.~Sanz-Cillero, F.~Zuo,
\newblock JHEP \textbf{1306} (2013) 020
  [\href{http://arxiv.org/abs/1304.3618}{arXiv:1304.3618}].

\bibitem{Skyrme:1961vq}
T.~H.~R.~Skyrme,
\newblock Proc.~Roy.~Soc.~Lond. \textbf{A260} (1961) 127.

\bibitem{Skyrme:1961vr}
T.~H.~R.~Skyrme,
\newblock Proc.~Roy.~Soc.~Lond. \textbf{A262} (1961) 237.

\bibitem{Skyrme:1962vh}
T.~H.~R.~Skyrme,
\newblock Nucl.~Phys. \textbf{31} (1962) 556.

%\cite{'tHooft:1973jz}
\bibitem{'tHooft:1973jz}
  G.~'t Hooft,
  %``A Planar Diagram Theory for Strong Interactions,''
  \newblock Nucl.~Phys. \textbf{B72} (1974) 461.
 % Nucl.\ Phys.\ B {\bf 72}, 461 (1974).
  %%CITATION = NUPHA,B72,461;%%
  %3442 citations counted in INSPIRE as of 26 Jun 2013


\bibitem{Witten:1979kh}
E.~Witten,
\newblock Nucl.~Phys. \textbf{B160} (1979) 57.

\bibitem{Witten:1998xy}
E.~Witten,
\newblock JHEP \textbf{07} (1998) 006
  [\href{http://arxiv.org/abs/hep-th/9805112}{hep-th/9805112}].

\bibitem{Hata:2007mb}
H.~Hata, T.~Sakai, S.~Sugimoto,  S.~Yamato,
\newblock Prog.~Theor.~Phys. \textbf{117} (2007) 1157
  [\href{http://arxiv.org/abs/hep-th/0701280}{hep-th/0701280}].

\bibitem{Hashimoto:2008zw}
K.~Hashimoto, T.~Sakai,  S.~Sugimoto,
\newblock Prog.~Theor.~Phys. \textbf{120} (2008) 1093
  [\href{http://arxiv.org/abs/0806.3122}{arXiv:0806.3122}].

\bibitem{Pomarol:2008aa}
A.~Pomarol,  A.~Wulzer,
\newblock Nucl. Phys. \textbf{B809} (2009) 347
  [\href{http://arxiv.org/abs/0807.0316}{arXiv:0807.0316}].


\bibitem{Panico:2008it}
G.~Panico,  A.~Wulzer,
\newblock Nucl. Phys. \textbf{A825} (2009) 91
  [\href{http://arxiv.org/abs/0811.2211}{arXiv:0811.2211}],
  %  \bibitem{Domenech:2010aq}
  O.~Domenech, G.~Panico and A.~Wulzer,
  %``Massive Pions, Anomalies and Baryons in Holographic QCD,''
 \newblock Nucl. Phys. \textbf{A853}, 97(2011)
 [\href{http://arxiv.org/abs/1009.0711}{arXiv:1009.0711}].


\bibitem{Adkins:1983ya}
G.~S.~Adkins, C.~R. ~Nappi,  E.~Witten,
\newblock Nucl. Phys. \textbf{B228} (1983) 552.

\bibitem{Cherman:2009gb}
A.~Cherman, T.~D. ~Cohen,  M.~Nielsen,
\newblock Phys. Rev. Lett. \textbf{103} (2009) 022001
  [\href{http://arxiv.org/abs/0903.2662}{arXiv:0903.2662}].

\bibitem{Becciolini:2009fu}
D.~Becciolini, M.~Redi,  A.~Wulzer,
\newblock JHEP \textbf{01} (2010) 074
  [\href{http://arxiv.org/abs/0906.4562}{arXiv:0906.4562}].

%\bibitem{Cappiello:2010uy}
%Luigi Cappiello, Oscar Cata, and Giancarlo D'Ambrosio.
%\newblock Phys. Rev. \textbf{D83} (2011): 093006
%  [\href{http://arxiv.org/abs/1009.1161}{arXiv: 1009.1161}].

\bibitem{Cherman:2011ve}
A.~Cherman,  T.~Ishii,
\newblock Phys. Rev. \textbf{D86} (2012) 045011
  [\href{http://arxiv.org/abs/1109.4665}{arXiv:1109.4665}].

\bibitem{Zuo:2011vh}
\newblock F.~Zuo, F.~K.~Guo, T.~Huang (2011),
  [\href{http://arxiv.org/abs/1111.5235}{arXiv:1111.5235}].

\bibitem{Adkins:1984cf}
G.~S.~ Adkins, C.~R. ~Nappi,
\newblock Nucl. Phys. \textbf{B249} (1985) 507.

\bibitem{Hata:2008xc}
H.~Hata, M.~Murata,  S.~Yamato,
\newblock Phys. Rev. \textbf{D78} (2008) 086006
  [\href{http://arxiv.org/abs/0803.0180}{arXiv:0803.0180}].

\bibitem{Bijnens:1999sh}
J.~Bijnens, G.~Colangelo,  G.~Ecker,
\newblock JHEP \textbf{02} (1999) 020
  [\href{http://arxiv.org/abs/hep-ph/9902437}{hep-ph/9902437}].

\bibitem{Bijnens:1999hw}
J.~Bijnens, G.~Colangelo,  G.~Ecker,
\newblock Annals Phys. \textbf{280} (2000) 100
  [\href{http://arxiv.org/abs/hep-ph/9907333}{hep-ph/9907333}].

\bibitem{Ecker:1988te}
G.~Ecker, J.~Gasser, A.~Pich,  E.~de~Rafael,
\newblock Nucl. Phys. \textbf{B321} (1989) 311.

\bibitem{Ecker:1989yg}
G.~Ecker, J.~Gasser, H.~Leutwyler, A.~Pich,  E.~de~Rafael,
\newblock Phys. Lett. \textbf{B223} (1989) 425.

%\cite{Braaten:1986md}
\bibitem{Braaten:1986md}
  E.~Braaten, S.~-M.~Tse and C.~Willcox,
  %``Electroweak Form-factors Of The Skyrmion,''
\newblock  Phys. Rev. \textbf{D34} (1986) 1482.

\bibitem{Atiyah:1989dq}
M.~F. Atiyah,  N.~S. Manton,
\newblock Phys. Lett. \textbf{B222} (1989) 438.

\bibitem{Witten:1976ck}
E.~Witten,
\newblock Phys. Rev. Lett. \textbf{38} (1977) 121.

\bibitem{Cohen:2012wm}
T.~D.~Cohen,  V.~Krejcirik,
\newblock Phys.Rev. \textbf{C85} (2012) 035205
  [\href{http://arxiv.org/abs/1201.5389}{arXiv:1201.5389}].

\bibitem{Adkins1987}
G.~S.~Adkins,
\newblock {\em Chiral solitons},
\newblock World Scientific, Singapore, 1987.

\end{thebibliography}
%\nocite{*}
\newpage

\end{document}